\documentclass[lettersize,journal]{IEEEtran}
\usepackage{array}
\usepackage[caption=false,font=normalsize,labelfont=sf,textfont=sf]{subfig}
\usepackage{stfloats}
\usepackage{url}
\usepackage{verbatim}
\usepackage{cite}
\usepackage{tabularray}
\usepackage{amsmath,amssymb,amsfonts}
\usepackage{algorithm}
\usepackage{algpseudocode}
\usepackage{booktabs,threeparttable} 
\usepackage{multirow}
\usepackage{subcaption}
\usepackage{graphicx}
\usepackage{textcomp}
\usepackage{xcolor}
\usepackage{svg}
\usepackage[caption=false,font=normalsize,labelfont=sf,textfont=sf]{subfig}
\usepackage{colortbl}

\newcommand{\newtext}{\textcolor{black}}
\newcommand{\reformulatedtext}{\textcolor{black}}

\newcommand{\revision}[1]{\textcolor{black}}

\newcommand{\EatOneArg}[1]{}

\begin{document}

\title{PowerFlowMultiNet: Multigraph Neural Networks for Unbalanced Three-Phase Distribution Systems}

\author{Salah~GHAMIZI, Jun~CAO,~\IEEEmembership{Member,~IEEE}, Aoxiang~MA, Pedro RODRIGUEZ,~\IEEEmembership{Fellow,~IEEE}

}

\markboth{Journal of \LaTeX\ Class Files,~Vol.~14, No.~8, August~2021}%
{Shell \MakeLowercase{\textit{et al.}}: A Sample Article Using IEEEtran.cls for IEEE Journals}


\maketitle

\begin{abstract}
Efficiently solving unbalanced three-phase power flow in distribution grids is pivotal for grid analysis and simulation. There is a pressing need for scalable algorithms
capable of handling large-scale unbalanced power grids that can provide accurate and fast solutions. To address this, deep learning techniques, especially Graph Neural Networks (GNNs), have emerged. However, existing literature primarily focuses on balanced networks, leaving a critical gap in supporting unbalanced three-phase power grids.

This letter introduces PowerFlowMultiNet, a novel multigraph GNN framework explicitly designed for unbalanced three-phase power grids. The proposed approach models each phase separately in a multigraph representation, effectively capturing the inherent asymmetry in unbalanced grids. A graph embedding mechanism utilizing message passing is introduced to capture spatial dependencies within the power system network.

PowerFlowMultiNet outperforms traditional methods and other deep learning approaches in terms of accuracy and computational speed. Rigorous testing reveals significantly lower error rates and a notable increase in computational speed for large power networks compared to model-based methods.
\end{abstract}


\section{Introduction}
The unbalanced three-phase power flow problem (PF) stands as a critical imperative within the domain of analysis and simulation of distribution grids with huge amounts of distributed energy resources (DERs). Traditional model-based methods using linearized models such as DistFlow\cite{DistFlow} and its variants or data-driven aided linear models \cite{DataDriven} to solve the unbalanced three-phase OPF. Despite their convergence guarantees in providing accurate solutions for unbalanced three-phase OPF problems in small to moderately sized networks, the scalability of computational methods becomes a critical concern when applied to large power grids\cite{lin2023powerflownet}.

To speed up and provide accurate power flow solutions for large-scale power systems, researchers have turned to deep learning techniques as a promising alternative. Although most deep learning (DL) techniques traditionally employ fully connected neural networks (FCNN) \cite{hasan_survey_2020} or deep convolutional neural networks \cite{CNN}, there is a burgeoning body of literature exploring the applications of Graph Neural Networks (GNNs) to account for the underlying network topology and reduce the model's size \cite{liao_review_2022}. These approaches leverage the graph structure to capture the dependencies between different components of the power system, providing an accurate and efficient solution to OPF problems in large networks.

Although numerous approaches \cite{liao_review_2022} have shown success in handling balanced networks, there is a critical gap regarding support for unbalanced three-phase power grids—a prevalent scenario in most distribution grids. Unbalanced grids, characterized by asymmetry in the three phases, present unique challenges that traditional DL 
cannot adequately address.


\newtext{Two challenges remain: 
1) How to design a graph representation that faithfully supports 3-phase components of the network. Traditional GNN with all the phases merged in one vector limits the ability to learn representations of unbalanced grids and 2) how to train a GNN for unbalanced grids. Reliable learning on large graphs remains challenging, only partially addressed with techniques like Generalized Convolutions\cite{li2020deepergcn}.}


To address the existing gap in unbalanced three-phase power grids using GNNs, we propose in this letter a novel GNN, namely \emph{PowerFlowMultiNet}, which stands as the first multigraph GNN framework explicitly designed for unbalanced three-phase power grids. Our approach tackles the challenges of unbalanced systems by modeling each of the three phases separately in a multigraph representation. 


\newtext{A multigraph is defined as a set of nodes interconnected by several types of edges. With a sub-graph constructed for each edge type, multigraphs can be thought of as the composition of many graphs sharing a common node set. Multigraphs have been extensively used to model and analyse brain networks\cite{lim2019discordant} and hate speech spread \cite{butler2023convolutional}. To the best of our knowledge, we are the first to propose their utilization to model power grids, and to argue their superiority for unbalanced distribution grids.}


Our nuanced approach allows us to capture the inherent asymmetry in unbalanced grids more effectively. Additionally, we introduce a graph embedding mechanism that leverages message passing to capture spatial dependencies and relationships within the power system network.

Our proposed approach, PowerFlowMultiNet, demonstrates compelling effectiveness compared to traditional model-based and deep learning methods. Through rigorous testing and evaluation, PowerFlowMultiNet achieved significantly lower errors than other deep learning approaches while providing a notable increase in computational speed for large power networks compared to model-based methods.

In summary, our contributions in this letter are two-fold:

\begin{enumerate}
    \item An automated framework for converting unbalanced three-phase modelling of components from OpenDSS to PyTorch graph representations. The framework uniquely supports multiphase transformers, lines, and loads, providing a robust tool for unbalanced power grid analysis.
    \item The design and implementation of a novel training architecture for multigraph GNNs, specifically tailored for the estimation of unbalanced OPF. This architecture leverages the strengths of GNNs in handling complex network structures, making it particularly suitable for unbalanced distribution grids applications.
\end{enumerate}



\section{Multigraph Neural Networks}
\begin{figure}[!ht]
\centering
\vspace{-1.em}
\includegraphics[width=\linewidth]{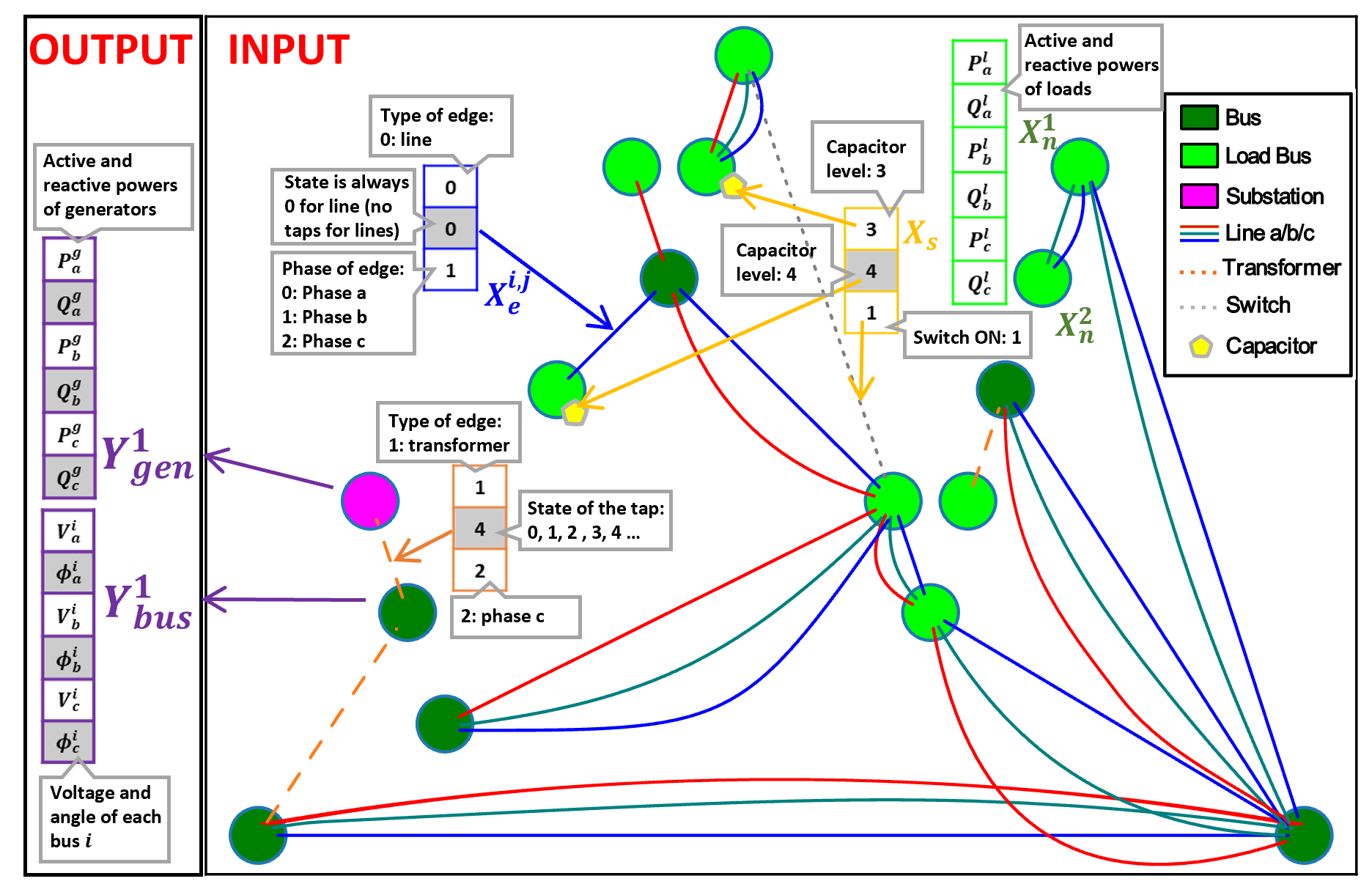}
\caption{\reformulatedtext{PowerFlowMultiNet graph embedding for IEEE 13-bus}}
\label{graph_embedding}
\vspace{-1.em}
\end{figure}

PowerFlowMultiNet framework includes two stages: (1) an unbalanced power grid to graph embedding, and (2) a powerflow learning with a novel GNN and training algorithm.

\subsection{Graph Embedding}

PowerFlowMultiNet uses PyDSS\footnote{https://nrel.github.io/PyDSS/index.html}, a wrapper for OpenDSS, to parse \emph{.dss} files and interact with its API. For example, given the IEEE Case 13 OpenDSS definition (Fig.\ref{graph_embedding}), we build an undirected multigraph representation $\mathcal{G}(\mathcal{N},\mathcal{E})$ of the topology of the power grid. The set of nodes of the graphs $\mathcal{N}$ consists of the substations (purple in Fig.\ref{graph_embedding}), loads (light green) and normal buses (dark green), and the set of edges of the graph $\mathcal{E}$ are the lines and transformers. To support unbalanced grids, each node stores its active power $P$ and its reactive power $Q$ for each phase a, b, and c as a feature $X_n\in \mathbb{R}^{N\times F_n}$. Two nodes are connected by one, two, or three edges depending on the number of phases of the connection. Each edge also stores in its features denoted as $X^e \in \mathbb{R}^{E\times F_e}$, the phase, the type (line or transformer), and the state of its optional tap. 

OpenDSS is used to estimate the unbalanced PF by updating the taps and capacitors' states accordingly. We update our graph features (load nodes P and Q, state of capacitors and transformers), and record the P and Q of the substation, \newtext{ and the buses voltage magnitude $V$ and angle $\phi$ } as node regression \reformulatedtext{outputs} (target features).


\revision{Hence, our model's inputs are the edges' features $X^e$ - describing the type of edge (0=line / 1=transformer), the state of the transformer's tap (0 if the edge is a line), and phase (0=a / 1=b / 2=c), highlighted in blue in Fig.\ref{graph_embedding}, and the loads' features $X^n$ - describing loads' active and reactive powers (the green features in Fig.\ref{graph_embedding}). The models' outputs are the active and reactive powers of the substations $Y^g$ and the voltage magnitude and angles of the buses $Y^b$ (in violet in Fig.\ref{graph_embedding})}. 

The features of the nodes and edges are processed in a TorchGeometric representation, their associated adjacency matrices $\mathcal{M}$ \revision{(one matrix per phase)} computed, and concatenated in a mini-batch pipeline. 
\subsection{GNN Training}

\begin{figure*}[!ht]
\centering
\vspace{-1.5em}
\includegraphics[trim={0 0.1cm 0 0.0cm},clip,width=1\linewidth]{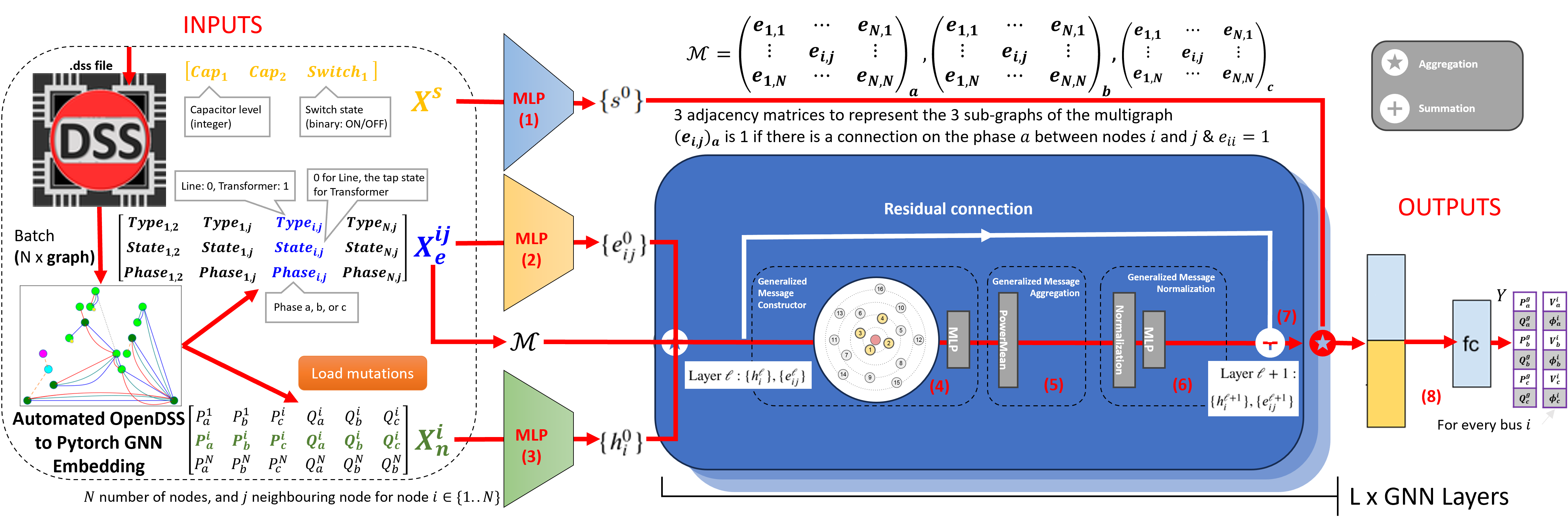}
\caption{Training of PowerFlowMultiNet: The embedding from Fig.1 is generated with OpenDSS, then concatenated into batches}
\label{graph_training}
\vspace{-1.5em}
\end{figure*}

We demonstrate the training in Fig.\ref{graph_training}. Given a batch of N sub-graphs\reformulatedtext{, at each layer $l$ of our GNN, the initial state of the capacitors and switches $s^{l=0}$ are encoded in (step 1 in Fig.\ref{graph_training}) by a multi-layer perceptron (MLP), the edge features $\mathbf{e}^{l=0}_{ij}$ between node $i$ and node $j$ are one-hot encoded the processed by an MLP (step 2). Similarly, the node features $\mathbf{h}^{l=0}_{ij}$ are normalized and then fed to an embedding MLP (step 3). Both edge embeddings $\mathbf{e}^{l}_{ij}$ and feature embeddings $\mathbf{h}^{l}_{ij}$ are concatenated. We refer to the concatenated vector as $h^{l}_v$}. 

Learning a graph representations usually learn latent features or representations for graphs using message passing (GCN)\cite{lin2023powerflownet}. 
The traditional GCN message passing operator $\mathcal{F}$ applied to node $i \in \mathcal{N}$ at the $l$-th layer is defined as follows:
\vspace{-0.5em}
\begin{align}
    \label{eq:mvu}
    &\mathbf{m}_{ij}^{(l)} = {\rho}^{(l)}(\mathbf{h}_{i}^{(l)}, \mathbf{h}_{j}^{(l)}, \mathbf{e}_{ij}^{(l)}), ~j \in \mathcal{N}(i) \\
    \label{eq:mv}
    &\mathbf{m}_{i}^{(l)} = {\zeta}^{(l)} ({\mathbf{m}_{ij}^{(l)}|u \in \mathcal{N}(i)}) \\
    \label{eq:hv+}
    &\mathbf{h}_{i}^{(l+1)} = {\phi}^{(l)}(\mathbf{h}_{i}^{(l)}, \mathbf{m}_{i}^{(l)}),
\end{align}
where ${\rho}^{(l)}, {\zeta}^{(l)}$, and ${\phi}^{(l)}$ are all differentiable functions for \emph{message construction}, \emph{message aggregation}, and \emph{feature update} in the $l$-th layer, respectively. For each node $i$, the message construction and aggregation takes into account all the nodes $j$ in its neighbor $\mathcal{N}(i)$. 
The node update function ${\phi}^{(l)}$ combines the features of the original node $\mathbf{h}_{i}^{(l)}$ and the aggregated message $\mathbf{m}_{i}^{(l)}$ to obtain the transformed features of the node $\mathbf{h}_{i}^{(l+1)}$.

In our approach, we use $L$ layers of type \emph{GEN}eralized \emph{CONV}olution \cite{li2020deepergcn}. A GENCONV layer extends the previous operators with a generalized message construction mechanism (step 4), a generalized message aggregator (step 5) mechanism, and a message normalization mechanism (step 6).

\paragraph{Message Construction}

We define the message construction function ${\rho}^{(l)}$ as follows:
\vspace{-0.5em}
\begin{align}
    \label{eq:gen_mc}
    \mathbf{m}_{ij}^{(l)} &= {\rho}^{(l)}(\mathbf{h}_{i}^{(l)}, \mathbf{h}_{j}^{(l)}, {\mathbf{e}_{ij}}^{(l)}) \nonumber\\ &=\text{ReLU}(\mathbf{h}_{j}^{(l)}+{\mathbf{e}_{ij}}^{(l)})+\epsilon, ~j \in \mathcal{N}(i),
\end{align}
where the $\text{ReLU}(\cdot)$ function is a rectified linear unit that bounds the output values to be greater than or equal to zero, $\epsilon$ is a small positive constant chosen as $10^{-7}$.

\paragraph{Message Aggregation}

We use as a generalized message aggregator operator a generalized mean-max aggregation function, denoted $\text{PowerMean\_Agg}_{p}(\cdot)$. This aggregation is valid, as the above message construction method always outputs positive values. The message aggregation ${\zeta}^{(l)}(\cdot)$ becomes:
\vspace{-1.5em}
\begin{align}
    \label{eq:gen_ma}
\mathbf{m}_{i}^{(l)} & = {\zeta}^{(l)} ({\mathbf{m}_{ij}^{(l)}|u \in \mathcal{N}(i)})  \nonumber\\  &= \text{PowerMean\_Agg}_{p}({\mathbf{m}_{ij}^{(l)}|u \in \mathcal{N}(i)}) \nonumber\\ &= (\frac{1}{\left|\mathcal{N}(i)\right|} \sum_{j \in \mathcal{N}(i)} \mathbf{m}_{ij}^{p})^{1/p},
\end{align}
where $p$ is a nonzero, continuous variable denoting the $q$-th power. It is a learnable parameter initialized at 1.

\paragraph{Message Normalization}

The main idea of this mechanism \cite{li2020deepergcn} is to normalize the features of the aggregated message $\mathbf{m}_{i}^{(l)} \in \mathbb{R}^{D}$ by combining them with the other features during the update phase. The update function becomes as follows:
\vspace{-0.5em}
\begin{align}
    \label{eq:mn}
    \mathbf{h}_{i}^{(l+1)} & = {\phi}^{(l)}(\mathbf{h}_{i}^{(l)}, \mathbf{m}_{i}^{(l)}) \nonumber \\ & = \text{MLP}(\mathbf{h}_{i}^{(l)} + s \cdot \lVert\mathbf{h}_{i}^{(l)}\rVert_2 \cdot \frac{\mathbf{m}_{i}^{(l)}}{\lVert\mathbf{m}_{i}^{(l)}\rVert_2} ),
\end{align}
where $\text{MLP}(\cdot)$ is a multi-layer perceptron and $s$ is a learnable scaling factor.
Residual connections allow the training of very deep GCN architectures without gradient vanish, and they also improve the stability of the learning. We implement the residual connections between the input of MsgConv  and the final addition of the layer (green arrow, step 7 in Fig.\ref{graph_training}).

\paragraph{State Concatenation}

After processing the graph embedding by $L\times$ GENCONV layers, the embedding is concatenated with the state embedding (obtained in step 1). The concatenated vector is then processed by one fully connected layer to output the active and reactive powers of the substation for each phase a, b, and c (step 8).

\begin{table*}[!hb]
\centering
\small
\caption{Results of PowerFlowMultiNet, GNN, and FCNN with real loads and perturbation of $\pm 10\%$ and $\pm 50\%$.} 
\label{results_uniform}
\resizebox{0.85\textwidth}{!}{
\begin{threeparttable}
    
    \small
\begin{tabular*}{1.2\linewidth}{ l l l | l l l l || l l }
    \toprule

Load       & Case    & Approach & $P_{MEAN}$\tiny{($STD$)}              & $Q_{MEAN}$\tiny{($STD$)}                      & $\phi_{MEAN}$\tiny{($STD$)}          & $V_{MEAN}$\tiny{($STD$)}   & Time$_{MEAN}$\tiny{($STD$)} & Time(Solver)$_{MEAN}$\tiny{($STD$)}           \\\midrule
$\pm 10\%$ & 13-Bus  & OURS     & \cellcolor[gray]{0.9} $2.47e^{-5}$\tiny{($2.35e^{-5}$)}   & $\cellcolor[gray]{0.9}2.61e^{-5}$\tiny{($2.11e^{-5}$)}           & $\cellcolor[gray]{0.9}5.72e^{-6}$\tiny{($2.4e^{-6}$)}  & $6.42e^{-6}$\tiny{($2.02e^{-6}$)} & $8.26e^{-6}$\tiny{($6.36e^{-7}$)}& $1.3e^{-4}$\tiny{($1.6e^{-4}$)}\\
           &         & GNN \cite{liao_review_2022}      & $4.35e^{-5}$\tiny{($4e^{-5}$)}      & $6.42e^{-5}$\tiny{($6.62e^{-5}$)}           & $1.15e^{-5}$\tiny{($6.26e^{-6}$)} & $\cellcolor[gray]{0.9}4.62e^{-6}$\tiny{($3.09e^{-6}$)}  & $1.05e^{-5}$\tiny{($1.2e^{-6}$)} & $\vert$\\
           &         & FCNN \cite{hasan_survey_2020}     & $2.23e^{-4}$\tiny{($2.24e^{-4}$)}   & $2.34e^{-4}$\tiny{($2.37e^{-4}$)}           & $7.43e^{-6}$\tiny{($4.91e^{64}$)} & $8.62e^{-6}$\tiny{($5.93e^{-6}$)} & $4.27e^{-6}$\tiny{($2.35e^{-7}$)} & $\vert$\\
           \cline{2-9}
           & 123-Bus & OURS     & \cellcolor[gray]{0.9}$1.26e^{-5}$\tiny{($1.5e^{-5}$)}    & \cellcolor[gray]{0.9}$1.09e^{-5}$\tiny{($1.08e^{-5}$)}           & \cellcolor[gray]{0.9}$2.02e^{-6}$\tiny{($2.55e^{-6}$)} & $1.18e^{-6}$\tiny{($6.2e^{-7}$)}   & $3.4e^{-5}$\tiny{($8.6e^{-6}$)}& $4.5e^{-4}$\tiny{($1.9e^{-4}$)}\\
           &         & GNN \cite{liao_review_2022}      & $2.41e^{-5}$\tiny{($2.24e^{-5}$)}   & $1.18e^{-5}$\tiny{($1.65e^{-5}$)}           & $3.69e^{-6}$\tiny{($1.99e^{-6}$)} & $\cellcolor[gray]{0.9} 9.92e^{-7}$\tiny{($6.11e^{-7}$)}  & $2.94e^{-5}$\tiny{($6.79e^{-6}$)} & $\vert$\\
           &         & FCNN \cite{hasan_survey_2020}     & $4.04e^{-5}$\tiny{($3.49e^{-5}$)}   & $2.88e^{-5}$\tiny{($2.52e^{-5}$)}           & $1.16e^{-5}$\tiny{($9.03e^{-6}$)} & $1.65e^{-6}$\tiny{($9.36e^{-7}$)}  & $2.11e^{-5}$\tiny{($4.04e^{-5}$)} & $\vert$\\
           \cline{2-9}
           & 906-Bus & OURS     & \cellcolor[gray]{0.9}$9.39e^{-8}$\tiny{($1.06e^{-7}$)}  & \cellcolor[gray]{0.9}$1.29e^{-7}$\tiny{($6.18e^{-8}$)}          & \cellcolor[gray]{0.9}$5.03e^{-6}$\tiny{($2.93e^{-6}$)} & \cellcolor[gray]{0.9}$4.31e^{-6}$\tiny{($2.39e^{-6}$)}  & $8.46e^{-5}$\tiny{($1.36e^{-5}$)}& $2.43e^{-3}$\tiny{($7.2e^{-4}$)}\\
           &         & GNN \cite{liao_review_2022}      & $4.57e^{-7}$\tiny{($4.37e^{-7}$)}  & $4.03e^{-7}$\tiny{($4.32e^{-7}$)}          & $2.28e^{-5}$\tiny{($4.20e^{-5}$)} & $8.45e^{-6}$\tiny{($6.64e^{-6}$)}  & $9.75e^{-6}$\tiny{($3.74e^{-7}$)} & $\vert$\\
           &         & FCNN \cite{hasan_survey_2020}     & $4.40e^{-7}$\tiny{($4.36e^{-7}$)} & $3.55e^{-7}$\tiny{($3.78e^{-7}$)} & $1.67e^{-5}$\tiny{($1.31e^{-6}$)} & $8.35e^{-6}$\tiny{($6.63e^{-6}$)} & $1.06e^{-6}$\tiny{($1.27e^{-7}$)} & $\vert$\\ \midrule
$\pm 50\%$ & 13-Bus  & OURS     & \cellcolor[gray]{0.9}$2.02e^{-4}$\tiny{($1.93e^{-4}$)}   & \cellcolor[gray]{0.9}$3.07e^{-4}$\tiny{($2.66e^{-4}$)}           & \cellcolor[gray]{0.9}$1.56e^{-5}$\tiny{($1.4e^{-5}$)}  & \cellcolor[gray]{0.9}$1.42e^{-5}$\tiny{($6.93e^{-6}$)}  & $1.01e^{-5}$\tiny{($6.49e^{-7}$)}& $1.9e^{-4}$\tiny{($1.7e^{-4}$)}\\
           &         & GNN \cite{liao_review_2022}      & $6.97e^{-4}$\tiny{($7.4e^{-4}$)}    & $5.91e^{-4}$\tiny{($6.13e^{-4}$)}           & $2.58e^{-5}$\tiny{($1.28e^{-5}$)} & $4.75e^{-5}$\tiny{($2.83e^{-5}$)}  & $1.06e^{-5}$\tiny{($1.29e^{-6}$)} & $\vert$\\
           &         & FCNN \cite{hasan_survey_2020}     & $2.27e^{-3}$\tiny{($2.56e^{-3}$)}   & $3.16e^{-3}$\tiny{($3.7e^{-3}$)}            & $1.02e^{-4}$\tiny{($6.03e^{-5}$)} & $1.16e^{-4}$\tiny{($7.16e^{-5}$)}  & $7.25e^{-6}$\tiny{($6.56e^{-6}$)} & $\vert$\\
           \cline{2-9}
           & 123-Bus & OURS     & \cellcolor[gray]{0.9}$6.28e^{-5}$\tiny{($5.7e^{-5}$)}    & \cellcolor[gray]{0.9}$4.68e^{-5}$\tiny{($3.76e^{-5}$)}            & \cellcolor[gray]{0.9}$1.07e^{-5}$\tiny{($8.94e^{-6}$)}  & \cellcolor[gray]{0.9}$5.27e^{-6}$\tiny{($2.73e^{-6}$)}  & $2.28e^{-5}$\tiny{($3.36e^{-7}$)}& $7.3e^{-4}$\tiny{($3.7e^{-4}$)}\\
           &         & GNN \cite{liao_review_2022}      & $3.75e^{-4}$\tiny{($3.63e^{-4}$)}   & $2.92e^{-4}$\tiny{($2.81e^{-4}$)}           & $1.44e^{-5}$\tiny{($8.62e^{-6}$)} & $1.61e^{-5}$\tiny{($8.84e^{-6}$)}  & $2.11e^{-5}$\tiny{($2.87e^{-7}$)} & $\vert$\\
           &         & FCNN \cite{hasan_survey_2020}     & $6.31e^{-4}$\tiny{($5.91e^{-4}$)}   & $2.92e^{-4}$\tiny{($5.03e^{-4}$)}           & $1.88e^{-5}$\tiny{($1.09e^{-5}$)} & $1.84e^{-5}$\tiny{($1.31e^{-5}$)} &  $7.76e^{-6}$\tiny{($2.26e^{-6}$)} & $\vert$\\
           \cline{2-9}
           & 906-Bus & OURS     & \cellcolor[gray]{0.9}$6.28e^{-7}$\tiny{($1.19e^{-7}$)}  & \cellcolor[gray]{0.9}$6.16e^{-8}$\tiny{($6.46e^{-8}$)}          & \cellcolor[gray]{0.9}$1.52e^{-5}$\tiny{($1.08e^{-5}$)} & \cellcolor[gray]{0.9}$2.2e^{-5}$\tiny{($1.47e^{-5}$)}  &  $6.68e^{-5}$\tiny{($2.11e^{-7}$)}& $2.53e^{-3}$\tiny{($6.1e^{-4}$)}\\
           &         & GNN \cite{liao_review_2022}      & $9.01e^{-6}$\tiny{($8.39e^{-6}$)}  & $7.43e^{-6}$\tiny{($7.5e^{-6}$)}          & $5.7e^{-5}$\tiny{($4.03e^{-5}$)} & $1.98e^{-4}$\tiny{($1.47e^{-4}$)}  & $9.51e^{-6}$\tiny{($1.55e^{-17}$)} & $\vert$\\
           &         & FCNN \cite{hasan_survey_2020}     & $1.07e^{-5}$\tiny{($1.03e^{-5}$)} & $8.42e^{-6}$\tiny{($8.07e^{-6}$)} & $4.15e^{-5}$\tiny{($3e^{-5}$)} & $2.03e^{-4}$\tiny{($1.51e^{-4}$)} & $8.90e^{-7}$\tiny{($9.04e^{-8}$)} & $\vert$\\ \midrule
Timeseries & 13-Bus  & OURS     & \cellcolor[gray]{0.9}$5.24e^{-4}$\tiny{($7.87e^{-4}$)}   & \cellcolor[gray]{0.9}$4.21e^{-4}$\tiny{($4.76e^{-4}$)}           & \cellcolor[gray]{0.9}$4.07e^{-5}$\tiny{($2.89e^{-5}$)} & \cellcolor[gray]{0.9}$2.66e^{-5}$\tiny{($2.28e^{-5}$)}  & $1.14e^{-5}$\tiny{($9.67e^{-7}$)}& $1.6e^{-4}$\tiny{($1e^{-4}$)}\\
           &         & GNN \cite{liao_review_2022}      & $1.19e^{-3}$\tiny{($1.53e^{-3}$)}   & $9.45e^{-3}$\tiny{($1.3e^{-3}$)}            & $4.47e^{-5}$\tiny{($3.51e^{-5}$)} & $6.95e^{-5}$\tiny{($6.78e^{-5}$)}  & $1.08e^{-5}$\tiny{($1.25e^{-6}$)} & $\vert$\\
           &         & FCNN \cite{hasan_survey_2020}     & $3.76e^{-3}$\tiny{($7.2e^{-3}$)}    & $4.32e^{-3}$\tiny{($8.21e^{-3}$)}           & $1.8e^{-4}$\tiny{($1.24e^{-4}$)}  & $1.71e^{-4}$\tiny{($1.84e^{-4}$)}  & $5.2e^{-5}$\tiny{($6.79e^{-5}$)} & $\vert$\\
           \cline{2-9}
           & 123-Bus & OURS     & \cellcolor[gray]{0.9}$1.09e^{-3}$\tiny{($1.16e^{-3}$)}   & \cellcolor[gray]{0.9}$8.44e^{-4}$\tiny{($8.4e^{-4}$)}            & \cellcolor[gray]{0.9}$5.22e^{-5}$\tiny{($2.65e^{-5}$)} & \cellcolor[gray]{0.9}$4.64e^{-5}$\tiny{($4.06e^{-5}$)}  & $2.76e^{-5}$\tiny{($3.51e^{-7}$)}& $1.03e^{-3}$\tiny{($5.6e^{-4}$)}\\
           &         & GNN \cite{liao_review_2022}      & $1.71e^{-3}$\tiny{($2.10e^{-3}$)}   & $1.56e^{-3}$\tiny{($2.44e^{-3}$)}           & $6.33e^{-5}$\tiny{($4.59e^{-5}$)} & $5.72e^{-5}$\tiny{($6.48e^{-5}$)}  & $2.12e^{-5}$\tiny{($3.51e^{-7}$)} &$\vert$\\
           &         & FCNN \cite{hasan_survey_2020}     & $3.68e^{-3}$\tiny{($5.82e^{-3}$)}   & $3.76e^{-3}$\tiny{($6.57e^{-3}$)}           & $8.21e^{-5}$\tiny{($5.94e^{-5}$)} & $8.42e^{-5}$\tiny{($1.15e^{-4}$)}  & $7.82e^{-6}$\tiny{($2.26e^{-6}$)} &$\vert$\\
           \cline{2-9}
           & 906-Bus & OURS     & \cellcolor[gray]{0.9}$2.15e^{-5}$\tiny{($2.25e^{-5}$)}  & \cellcolor[gray]{0.9}$1.15e^{-5}$\tiny{($1.2e^{-5}$)}           & \cellcolor[gray]{0.9}$4.58e^{-5}$\tiny{($2.05e^{-5}$)} & \cellcolor[gray]{0.9}$1.83e^{-4}$\tiny{($1.43e^{-4}$)}  & $6.66e^{-5}$\tiny{($2.37e^{-7}$)}& $2.46e^{-3}$\tiny{($5.6e^{-4}$)}\\
           &         & GNN \cite{liao_review_2022}      & $1.18e^{-4}$\tiny{($1.04e^{-4}$)}  & $7.7e^{-5}$\tiny{($7.83e^{-5}$)}          & $2.02e^{-4}$\tiny{($1.75e^{-4}$)} & $1.21e^{-3}$\tiny{($9.24e^{-4}$)}  & $9.48e^{-6}$\tiny{($1.4e^{-7}$)} &$\vert$\\
           &         & FCNN \cite{hasan_survey_2020}     & $4.2e^{-4}$\tiny{($4.05e^{-4}$)}  & $2.54e^{-4}$\tiny{($2.65e^{-4}$)}          & $2.11e^{-3}$\tiny{($2.7e^{-3}$)} & $2.32e^{-3}$\tiny{($1.56e^{-3}$)}  & $9.39e^{-7}$\tiny{($8.45e^{-8}$)} &$\vert$
\\             

\bottomrule
\end{tabular*}
\vspace{-1em}
\end{threeparttable}
}
\end{table*}

\section{Empirical Study}

\paragraph{Use cases}
We evaluate two commonly used unbalanced topologies: IEEE 13, 123-Bus and IEEE European LV network (906-Bus) systems. The hardware environment is a Nvidia V100-32GB Graphic Card with 2.2 GHz core clock.

\paragraph{Simulation tool} We use OpenDSS 
for topology validation and power flow simulation as ground truth. 
For each case, we mutate the network by varying the loads to generate the training and validation set. We use 8k training mutants and 2k validation mutants. We evaluate loads following a uniform distribution, with $\pm 10\%$, $\pm 50\%$ variations, and following real-world time series from London Kaggle Timeseries\footnote{https://www.kaggle.com/datasets/jeanmidev/smart-meters-in-london}.

\paragraph{Training and optimization}

We train all the models for 1000 epochs and batch size of 128 with MSE loss, Adam optimizer, and a multistep learning rate starting at 0.001.

\paragraph{Metrics}

We report the mean and standard deviation of two metrics: \reformulatedtext{the Normalized Squared Error to the active powers $P_{MEAN(STD)}$ and reactive powers $Q_{MEAN(STD)}$ of the substations, and to the voltage magnitude $V_{MEAN(STD)}$ and angle $\phi_{MEAN(STD)}$ of the buses.} The ground truth are obtained with the power flow simulation (OpenDSS). 
We also report the execution time (seconds) on the validation graphs, by comparing the forward operation of the deep learning model with the solve operation of OpenDSS. 

\paragraph{Results}

\reformulatedtext{We summarize the performance of our approach in Table \ref{results_uniform}. The gray cells indicate the lowest errors per scenario. 
PowerFlowMultiNet is always more precise than the FCNN and GNN models (except for voltage magnitude with small grids and 10\% load perturbation, where GNN achieves as low errors as our approach), while having execution costs similar to simpler approaches.
Overall, our approach achieves very low error compared to the OpenDSS ground truth.}

\reformulatedtext{Next, we evaluate the efficiency of our approach. Our approach is up to 15 times faster than Model-based on a 13-bus grid (e.g., for 10\% load variation), up to 37 times faster on 123-bus grid (e.g., for timeseries load variation), and up to 38 times faster on 906-bus (e.g., for 50\% load variation)}.

\reformulatedtext{Notably, our approach achieves stable execution time for all grid and a very low standard deviation, confirming its stability and scalability to larger grids. Our approach behaves similarly even for cases hard to optimize by model-based approaches. 
}


\section{Conclusion}
This paper is the first to tackle the unbalanced three-phase power flow problem as a multigraph neural network. Our framework, PowerFlowMultiNet, includes a novel graph embedding of unbalanced distribution grids and an efficient GNN training algorithm for power flow prediction. Our empirical study on 13, 123, and 906 bus distribution grids demonstrates its effectiveness compared to existing deep learning methods and its efficiency compared to model-based methods. We believe that our work paves the way for novel applications of GNN for distribution grid optimizations and operations.


{
\bibliography{bib/leap,bib/ml,bib/other}
}
\bibliographystyle{IEEEtran}

\newpage

\vfill

\end{document}